\documentclass[aps,prb,twocolumn,superscriptaddress,floatfix,longbibliography,citeautoscript]{revtex4-2}
\usepackage{relsize}
\usepackage{amsmath,amssymb}
\usepackage{bm} 
\usepackage{graphicx}
\usepackage{comment}
\usepackage[normalem]{ulem} 
\usepackage{enumitem}
\setlist{noitemsep,leftmargin=*,topsep=0pt,parsep=0pt}
\usepackage{xcolor}
\definecolor{lightgray}{gray}{0.6}
\definecolor{medgray}{gray}{0.4}
\definecolor{mRed}{RGB}{230, 0, 50}
\colorlet{newtextColor}{mRed}
\colorlet{deltextColor}{black}
\usepackage{hyperref}
\hypersetup{
colorlinks=true,
urlcolor= blue,
citecolor=blue,
linkcolor= blue,
}
\makeatletter
\renewcommand*{\fnum@figure}{\textbf{Fig.\,\thefigure}\,}
\renewcommand*{\@caption@fignum@sep}{\textbf{\,\textbar\,}}
\makeatother

\newif\ifptitle
\newif\ifpnumber
\newcounter{para}
\newcommand\ptitle[1]{\par\refstepcounter{para}
{\ifpnumber{\noindent\textcolor{lightgray}{\textbf{\thepara}}\indent}\fi}
{\ifptitle{\textbf{[{#1}]}}\fi}}

\newif\iftrackchanges

\usepackage{mdframed}
\newmdenv[
  linecolor={\iftrackchanges newtextColor\else white\fi},
  linewidth=2pt,
  topline=false,
  bottomline=false,
  rightline=false,
  skipabove=\topsep,
  skipbelow=\topsep,
  leftmargin=-12pt,
  innertopmargin=0pt,
  innerbottommargin=0pt
]{newtextblock}
\newcommand{\heng}{School of Engineering \& Applied Sciences, Harvard University, Cambridge, MA 02138, USA}
\newcommand{\hphys}{Department of Physics, Harvard University, Cambridge, MA 02138, USA}
\newcommand{\hchem}{Department of Chemistry, Harvard University, Cambridge, MA 02138, USA}
\newcommand{\Weizmann}{Department of Condensed Matter Physics, Weizmann Institute of Science, Rehovot, Israel}
\newcommand{\IL}{Department of Physics, University of Illinois Chicago, Chicago, IL 60607, USA}
\newcommand{\Row}{The Rowland Institute at Harvard, Harvard University, Cambridge, MA, 02138, USA}
\newcommand{\SL}{Department of Physics, Washington University in Saint Louis, Saint Louis, MO 63130, USA}
\newcommand{\PSU}{Department of Physics, The Pennsylvania State University, University Park, PA 16802, USA}

\newcommand{\mytitle}{Kondo Reshapes Multiple Orders in a $5f$ van der Waals Material}
\newcommand{\but}{$\beta$-UTe$_3$}

\newcommand{\Vs}{\ensuremath{V_{\mathrm{sample}}}}
\newcommand{\Is}{\ensuremath{I_{\mathrm{set}}}}
\newcommand{\Ve}{\ensuremath{V_{\mathrm{exc}}}}
\newcommand{\EF}{\ensuremath{E_{\mathrm{F}}}}

\makeatletter
\renewcommand\section{%
  \@startsection{section}{1}{\z@}%
  {1.5ex \@plus .5ex \@minus .2ex}
  {0.8ex \@plus .2ex}
  {\normalfont\large\bfseries\raggedright}%
}
\makeatother

\makeatletter
\renewcommand\subsection{%
  \@startsection{subsection}{2}{\z@}%
  {2ex \@plus .5ex \@minus .2ex}
  {0.01pt}
  {\normalfont\fontsize{11pt}{13pt}\bfseries\raggedright}%
}
\makeatother

\makeatletter
\renewcommand\subsubsection{%
  \@startsection{subsubsection}{3}{\z@}%
  {1.5ex \@plus .5ex \@minus .2ex}
  {0pt}
  {\normalfont\normalsize\bfseries\raggedright}%
}
\makeatother

\makeatletter
\def\@hangfrom@section#1#2#3{\@hangfrom{#1#2}#3}
\def\@hangfroms@section#1#2{#1#2}
\makeatother

\ptitlefalse
\pnumbertrue 
\trackchangestrue

\begin{document}

\title{\mytitle}
\author{Gal Tuvia}
\thanks{These authors contributed equally}
\affiliation{\hphys}
\author{Ruizhe Kang}
\thanks{These authors contributed equally}
\affiliation{\heng}
\author{Diana Golovanova}
\affiliation{\Weizmann}
\author{Yuqian Chen}
\affiliation{\hchem}
\author{Yidi Wang}
\affiliation{\hphys}
\author{Zeyu Ma}
\affiliation{\heng}
\author{Mengke Liu}
\affiliation{\hphys}
\author{Carly Grossman}
\affiliation{\Row}
\author{Suk Hyun Sung}
\affiliation{\Row}
\author{Justin Shotton}
\affiliation{\SL}
\author{Jiahui Zhu}
\affiliation{\SL}
\author{David Martinez}
\affiliation{\SL}
\author{Ismail El Baggari}
\affiliation{\Row}
\author{Binghai Yan}
\affiliation{\Weizmann}
\affiliation{\PSU}
\author{Dirk K. Morr}
\affiliation{\IL}
\author{Sheng Ran}
\affiliation{\SL}
\author{Jennifer E. Hoffman}
\email[]{jhoffman@physics.harvard.edu}
\affiliation{\hphys}
\affiliation{\heng}

\date{\today}

\begin{abstract}
Electron interactions can drive magnetism, superconductivity, and topology. However, the realization of these phases remains limited in van der Waals materials, and the full landscape of strong correlations remains uncharted in any context. While interactions between conduction electrons and localized spins yield a well-known competition between heavy fermions (Kondo hybridization) and magnetic order (RKKY exchange), such \textit{spin}-driven competition represents only part of the correlated electron phase diagram. Here we demonstrate that a heavy-fermion state can also compete with \textit{charge} order, such as the charge density wave (CDW) state typical in the van der Waals $4f$ rare-earth tritellurides (RTe$_3$). We exploit the spatially-extended $5f$ orbitals of \but\ to enhance Kondo hybridization compared to its isostructural RTe$_3$ cousins. Our scanning tunneling spectroscopy on \but\ shows Fano resonances characteristic of the heavy fermion state, while our quasiparticle interference imaging reveals the disappearance of Fermi-level nesting and the appearance of flat bands. We extend the tritelluride tight-binding model to include Kondo coupling and quantify the Fermi surface reconstruction. Consistent with the destruction of nesting, we observe no CDW in \but. Our expansion of the Kondo phase diagram beyond spin-mediated competition opens new possibilities for proximity-induced phase engineering in correlated van der Waals heterostructures.
\end{abstract}

\maketitle
\ptitle{Kondo vs RKKY vs other correlated phases}
Within the landscape of electron correlations, the coupling between localized magnetic moments and itinerant carriers generates a rich phase diagram. When the coupling is weak, the Ruderman–Kittel–Kasuya–Yosida (RKKY) interaction dominates, giving a magnetically ordered ground state~\cite{jensenOxford1991}; for stronger coupling, the Kondo interaction reconstructs the Fermi surface into heavy quasiparticle bands~\cite{colemanArxiv2015}. Tuning the Kondo–RKKY competition can drive quantum criticality~\cite{stewartRevModPhys2001,paschenNatRPhys2021} and give rise to functional correlated phases such as superconductivity~\cite{ranScience2019,jiaoNature2020,checkelskyNatRMatt2024}.
But this well-known Doniach phase diagram~\cite{Doniach1977} represents only part of the landscape of correlated electronic phases; because Kondo hybridization reconstructs the Fermi surface, it may compete with multiple instabilities.

\ptitle{Tritellurides as platform for Kondo}
To explore the broader competition between multiple correlated orders, we search for a system with localized spins whose order depends sensitively on its Fermi surface structure---conditions found in the
rare-earth tritellurides (RTe$_3$). The Te 5$p$ orbitals form quasi-one-dimensional conduction bands, yielding a strongly-nested Fermi surface and driving charge density wave (CDW) order~\cite{brouetPRB2008,brouetPRL2004,fangPRL2007,tomicPRB2009,yaoPRB2006}. The rare-earth sites host $4f$ moments that may hybridize with the itinerant electrons and modify the nested Fermi-surface instability. However, RTe$_3$ compounds exhibit antiferromagnetic order with N{\'e}el temperatures that scale with the de Gennes factor~\cite{ruPRB2008} of the rare earth element, behavior characteristic of RKKY interactions~\cite{jensenOxford1991}, leaving the Kondo regime of the phase diagram unexplored.

\begin{figure*}[t]
    \includegraphics[clip=true,width=\textwidth]{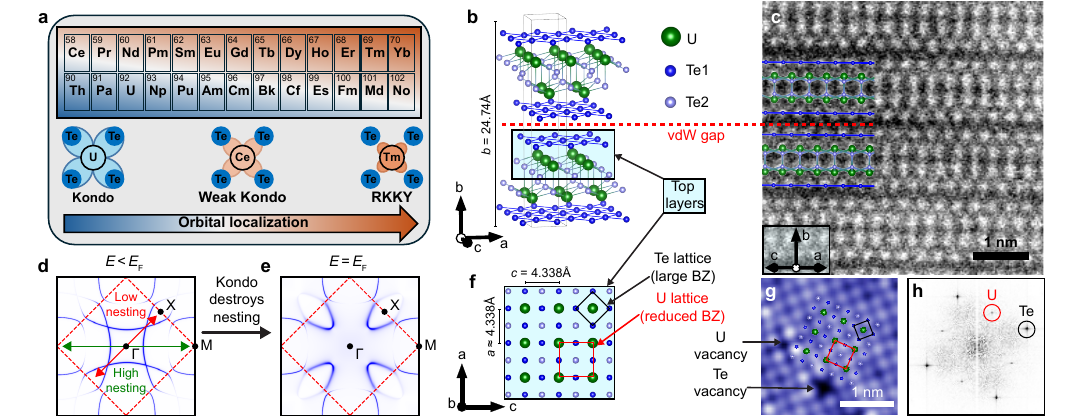}
    \caption{\textbf{Tritelluride overview and \but\ structure.} \textbf{a}, Evolution of the RKKY--Kondo crossover in the tritelluride series, governed by $f$-orbital localization. TmTe$_3$ exhibits localized $4f$ orbitals and RKKY-dominated interactions; CeTe$_3$ has more extended $4f$ orbitals and weak Kondo behavior; \but\ has even more extended 5$f$ orbitals and lies fully in the Kondo regime.
    \textbf{b}, Crystal structure of \but, showing the van der Waals (vdW) gap and the layers exposed by cleaving.
    \textbf{c}, Cross-sectional STEM image viewed along the $a$–$c$ plane with the $b$ axis vertical; the crystal structure is overlaid.
    \textbf{d},\textbf{e}, Tight-binding calculation of constant-energy contours in \but~(see Methods). The nested structure at energies below \EF\ (\textbf{d}) is destroyed at the Fermi surface (\textbf{e}) due to Kondo hybridization.
    \textbf{f}, Top view of the \but\ crystal structure, highlighting the Te and U sub-lattices.
    \textbf{g}, STM topograph with the lattice overlaid; a U vacancy and a Te vacancy are indicated.
    \textbf{h}, Fourier transform of (\textbf{g}), showing peaks from the Te and U periodicities (calculated from the larger field of view in Extended Data Fig.~\ref{fig:full_topo}a).
    STM topograph was measured at $T = 4.5$ K with sample bias $\Vs = -200$ mV and current setpoint $\Is = 100$ pA.
    }
    \label{fig:KS}
\end{figure*}

\ptitle{Driving Kondo in tritellurides}
To drive a system from the RKKY to the Kondo regime, it is useful to consider the Kmetko–Smith diagram, which links $f$-orbital localization to the electronic ground state~\cite{smithJLCM1983,colemanArxiv2015}. RKKY magnetic order is favored for more localized $f$ orbitals, while Kondo screening emerges as the orbitals become more extended and hybridize more strongly with itinerant electrons. Because the 4$f$ orbital size expands with decreasing atomic number, CeTe$_3$ is the most likely RTe$_3$ compound to exhibit Kondo physics (Fig.~\ref{fig:KS}a).  However, CeTe$_3$ exhibits only weak Kondo behavior,
where a heavy Fermi surface does not emerge~\cite{ruPRB2006,brouetPRB2008,trontlARxiv2025,ZengNewton2026}. To tune the system further into the Kondo regime, we move down the periodic table to the actinides, where the more spatially extended 5$f$ orbitals may tip the balance toward a heavy-fermion state. Recent work has shown that \but\ differs from the RTe$_3$ family, exhibiting ferromagnetism along with a large Sommerfeld coefficient indicative of a large electronic mass~\cite{thomasArxiv2025}. Yet the effect of Kondo hybridization on the Fermi surface and the resulting correlated phase diagram is unmapped.

\ptitle{Main findings}
Here, we show that the $5f$ tritelluride \but\ realizes the heavy-fermion state absent in the $4f$ tritellurides. Using scanning tunneling microscopy (STM), we observe a Fano resonance characteristic of Kondo hybridization. We image quasiparticle interference (QPI), revealing strong nesting away from the Fermi level \EF, whereas the Fermi surface itself is reconstructed into new flat bands. Our tight-binding model shows that the observed QPI modes are consistent with a reconstruction of the nested Fermi surface into heavy bands through Kondo coupling to the U sites. Unlike in the RTe$_3$ compounds, we find no evidence for CDW order in \but, in either STM topographs or in complementary scanning transmission electron microscopy (STEM) measurements. We thus show that Kondo-driven destruction of Fermi-surface nesting suppresses CDW formation and may also explain the recently-reported ferromagnetic order~\cite{thomasArxiv2025}, establishing the tritellurides as a platform where Kondo interactions can reshape a broad range of correlated electronic phenomena.

\ptitle{Crystal structure}
Like the rare-earth tritellurides, \but\ is a van der Waals (vdW) material that crystallizes stably in the orthorhombic space group $Cmcm$ (No.~63)~\cite{NoelJSSC1989,thomasArxiv2025,Sakai2025note}. It is composed of a UTe block surrounded by two Te layers (Fig.~\ref{fig:KS}b), separated by a vdW gap that can be seen in the STEM section image (Fig.~\ref{fig:KS}c). The in-plane lattice constants are nearly equal so it is approximated as pseudo-tetragonal ($a \approx c= 4.338$~\AA, $b = 24.74$~\AA)~\cite{NoelJSSC1989,thomasArxiv2025}. The in-plane unit cell area of each Te layer is half that of the crystallographic in-plane unit cell defined by the UTe block (Fig.~\ref{fig:KS}f). Because the itinerant electrons are largely confined to the Te layers, the Brillouin zone (BZ) is often considered for the Te layer alone and then folded in half when the UTe block is introduced as a perturbation to the periodic potential~\cite{voitScience2000,brouetPRL2004}.

\ptitle{Information from topography}
An STM topograph of the cleaved \but\ surface is shown in Fig.~\ref{fig:KS}g (cropped from the larger image in Extended Data Fig.~\ref{fig:full_topo}a), where both U and Te vacancies are visible. The Fourier transform of the full topograph, shown in Fig.~\ref{fig:KS}h, displays two sets of Bragg peaks: those marked in red correspond to the U sub-lattice associated with the reduced BZ, while those in black correspond to the Te sub-lattice. The Te peaks exhibit stronger Fourier intensity in this scan, indicating that the Te sub-lattice dominates the real-space topograph. However, the Bragg peak contrast varies with STM setpoint voltage and tip material; we attribute this variability to co-tunneling processes characteristic of Kondo-hybridized systems (see Extended Data Figs.~\ref{fig:topo_tips} and \ref{fig:fano_tips}).

\begin{figure}[t]
    \includegraphics[clip=true,width=\columnwidth]{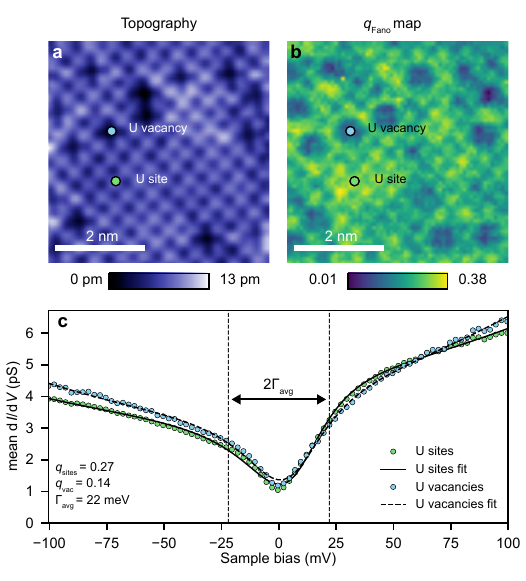}
    \caption{\textbf{\but\ defect imaging and Kondo spectroscopy.} \textbf{a}, STM topograph and \textbf{b}, Fano $q$ map, fitted from a d$I$/d$V$ map in the same area. A representative U site and U vacancy are marked in both panels; $q$ is high on U sites and suppressed on U vacancies. 
    \textbf{c}, d$I$/d$V$ spectra averaged over U sites (green) and U vacancies (cyan), and corresponding Fano line shape fits to equation~(\ref{eq:stsfit}). The fitted hybridization energy $2\Gamma$ is approximately the same for both curves, while $q$ varies.
    Measurements were performed at $T = 7.2$ K with $\Vs = 100$ mV and $\Is = 150$ pA; spectra were acquired using a zero-to-peak lock-in modulation amplitude of $\Ve = 4$ mV.
    }
    \label{fig:fano}
\end{figure}

\ptitle{Introduction to Fano}
Tunneling differential conductance spectroscopy (d$I$/d$V$) provides a direct probe of Kondo physics. When an electron tunnels into a Kondo-hybridized system, it simultaneously enters through the localized $f$ states and the itinerant conduction electrons. Interference between these paths produces a characteristic Fano line shape in the spectra~\cite{maltsevaPRL2009,FigginsPRL2010,morrRepProgPhys2016,SchmidtNature2010,AynajianPNAS2010,parkPRL2012,zhaoNanoLett2021,giannakisSciAdv2019,poseyNature2024,PirieScience2023,jiaoNature2020,zhang2013PRX,RoblerPNAS2014,zhangSciAdv2018,turkel2025NatPhys},
\begin{equation}
\frac{\mathrm{d}I}{\mathrm{d}V}\!(V)
= k\, \frac{\left(q + \displaystyle{\frac{V - E_{0}}{\Gamma}}\right)^{2}}
{1 + \left(\displaystyle{\frac{V - E_{0}}{\Gamma}}\right)^{2}}
+ a V^{2} + b V + c\,.
\label{eq:stsfit}
\end{equation}
Here $q$ is the asymmetry parameter, which varies monotonically with the tunneling ratio between the $f$ and itinerant states, but also depends on factors such as particle–hole asymmetry~\cite{morrRepProgPhys2016}. 
$\Gamma$ is the Kondo hybridization energy scale; 
$E_{0}$ is the resonance energy; 
$a$, $b$, and $c$ define a parabolic background; 
and $k$ sets the relative amplitude of the Fano contribution to the spectrum.

\ptitle{Spatial variation of Kondo hybridization}
We measure a grid of d$I$/d$V$ spectra over a $5 \times 5$~nm$^2$ region where U sites and U vacancies are clearly resolved (Fig.~\ref{fig:fano}a). We fit each spectrum to the Fano line shape in equation~(\ref{eq:stsfit}), producing the spatial map of the asymmetry parameter $q$ in Fig.~\ref{fig:fano}b. The map shows that $q$ peaks on U atoms, decreases between them, and is smallest at U vacancies. We compare the average spectra from all U sites and all vacancies in Fig.~\ref{fig:fano}c. Both averages give the same hybridization energy, $\Gamma = 22$~meV, while $q$ is markedly smaller on vacancies (0.14) than on U sites (0.27). This spatial variation reflects the physical meaning of $q$ as a tunneling ratio that increases monotonically as the tip moves closer to the source of $f$ electrons---U atoms. The magnitude of $q$ also varies between scans taken with different tip materials, reaching values as high as 1.3, whereas $\Gamma$ remains constant within 22–24~meV (Extended Data Fig.~\ref{fig:fano_tips}). Because the tunneling amplitudes into the two electronic channels depend on the orbital overlap with the tip, $q$ can change with tip material. In contrast, $\Gamma$ is the Kondo hybridization energy scale of \but, which should be independent of tip material, consistent with our observations. Similarly, experiments performed under different conditions reveal variations in the Fano lineshape parameters in URu$_2$Si$_2$~\cite{parkPRL2012,SchmidtNature2010,AynajianPNAS2010} and SmB$_6$~\cite{zhang2013PRX,PirieScience2023,RoblerPNAS2014}.

\begin{figure*}[t]
    \includegraphics[clip=true,width=\textwidth]{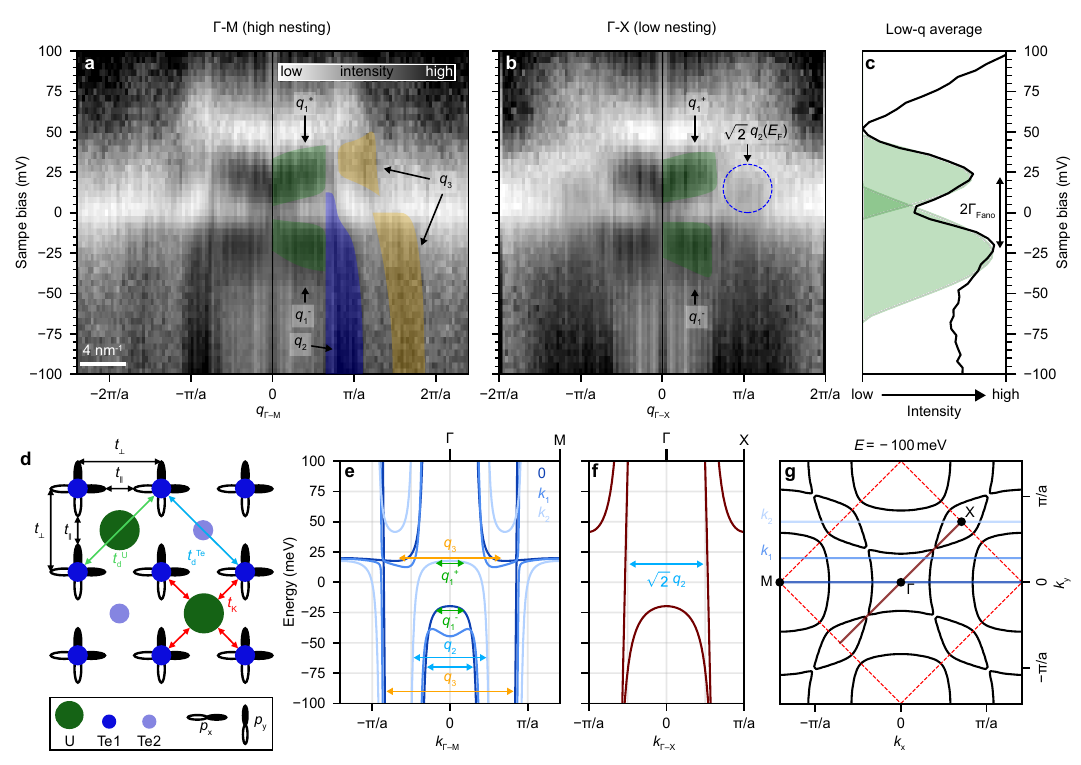}
    \caption{\textbf{Quasiparticle interference and tight-binding model of \but.}
\textbf{a},\textbf{b}, QPI dispersion along the $\Gamma$–M (\textbf{a}) and $\Gamma$–X (\textbf{b}) directions, extracted from the Fourier transforms of d$I$/d$V$ energy slices acquired in $25 \times 25$~nm$^2$ area (see Methods). The dominant scattering modes are marked as $q_1^{\pm}$ (green), $q_2$ (blue), and $q_3$ (orange), with colored guides to the eye. In (\textbf{b}), the blue dashed circle marks the $\Gamma$–X position corresponding to the $\Gamma$–M mode $q_2$, scaled by $\sqrt{2}$.
\textbf{c}, Energy-resolved QPI intensity obtained by integrating panels (\textbf{a}) and (\textbf{b}) over a low-$q$ window associated with the $q_1$ mode and then averaging the two directions (black curve). Green shading shows Gaussian fits to the two peaks. The hybridization gap is consistent with the energy scale $2\Gamma$ (black arrow) extracted from the Fano fits in Fig.~\ref{fig:fano}.
\textbf{d}, Tight-binding model of \but, with five coupling terms indicated: nearest-neighbor Te $5p$ orbital couplings ($t_{\parallel}$ and $t_{\perp}$, black); next-nearest-neighbor Te $5p$ orbital diagonal couplings across underlying U ($t_\mathrm{d}^\mathrm{U}$, green) and across underlying Te ($t_\mathrm{d}^\mathrm{U}$, cyan); and effective Kondo coupling between Te $5p$ and U $5f$ orbitals ($t_\mathrm{K}$, red). The terms capture the nested tritelluride band structure and the Kondo hybridization.
\textbf{e}, Calculated band structure cuts parallel to $\Gamma$–M, at the constant $k_y$ values marked in (\textbf{g}). \textbf{f},  Calculated band structure along $\Gamma$–X. QPI scattering modes $q_1^{\pm}$, $q_2$, and $q_3$ are labeled in colors corresponding to (\textbf{a}--\textbf{b}).
\textbf{g}, Constant-energy contour at $-100$ mV, below the Kondo energy range. The reduced Brillouin zone is outlined by the dashed red box; linecuts used in (\textbf{e}--\textbf{f}) are marked.
Measurements were performed at $T = 7.1$ K with $\Vs = 100$ mV, $\Is = 800$ pA, and $\Ve = 4$ mV. 
    }
    \label{fig:QPI}
\end{figure*}

\begin{figure*}[t]
    \includegraphics[clip=true,width=\textwidth]{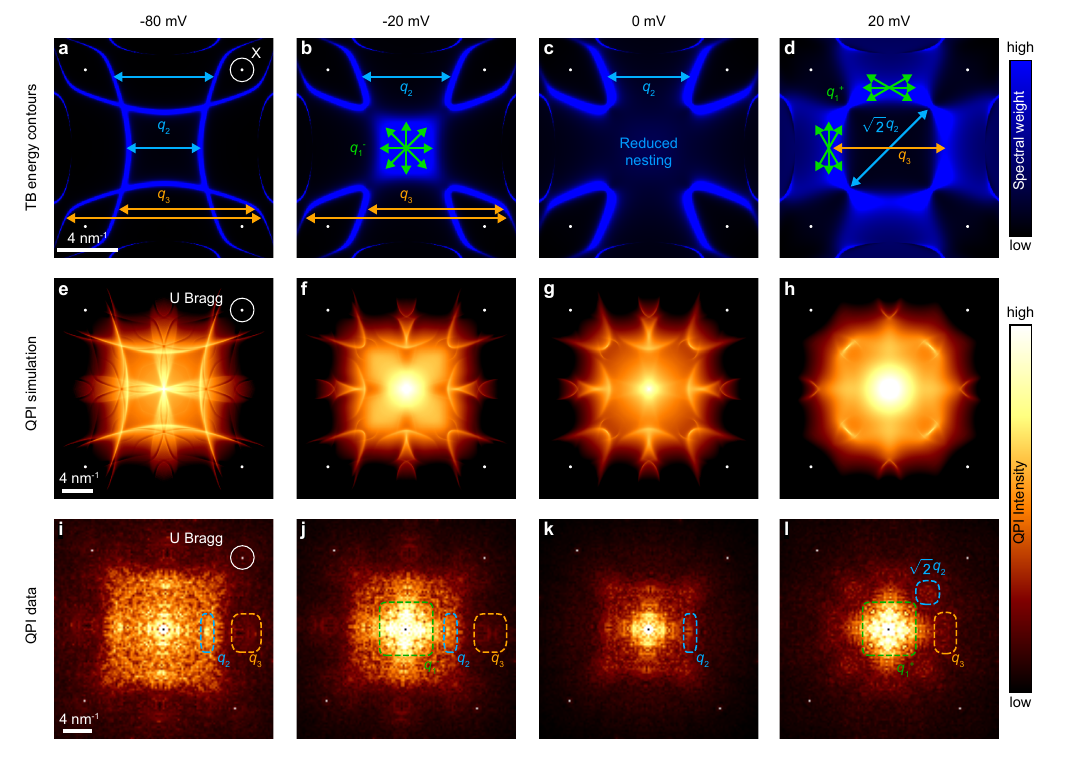}
    \caption{\textbf{Kondo-driven Destruction of Fermi-surface nesting in \but.}
\textbf{a}--\textbf{d}, Calculated spectral weight shown as constant-energy contours at selected energies, with the expected QPI modes indicated as colored arrows. Heavy bands responsible for the $q_1^{\pm}$ modes form at energies below (\textbf{b}) and above (\textbf{d}) the Fermi level. The central contour is destroyed at the Fermi level (\textbf{c}).
\textbf{e}--\textbf{h}, Simulated QPI heatmaps at the same energies (see Methods), with reduced-BZ Bragg peaks of the U lattice added for scale.
\textbf{i}--\textbf{l}, Experimental QPI constant-energy heatmaps at the same energies, with the main QPI modes labeled. The U sub-lattice Bragg peak is visible. Color scales are shared within every row; panels \textbf{e}--\textbf{h} are shown in log-scale to highlight lower intensity features.
}
    \label{fig:Energy_slices}
\end{figure*}

\ptitle{Introducing QPI}
To reveal the effect of Kondo hybridization on the Fermi surface, we image standing wave patterns formed by quasiparticle interference (QPI), whose energy layers can be Fourier transformed and analyzed to give the electronic band structure~\cite{crommieNature1993,avrahamAdvMat2018}. Fig.~\ref{fig:QPI}a,b shows the QPI dispersion along $\Gamma$–M and $\Gamma$–X, with the dominant scattering modes along each direction labeled as $q_1^{\pm}$, $q_2$, and $q_3$. At negative energies, below the Kondo regime, the QPI signal is strongly anisotropic: the scattering modes $q_2$ and $q_3$ appear along $\Gamma$–M, whereas modes along $\Gamma$–X are not as well-defined. This anisotropy reflects the nested band structure of the tritellurides, where many electronic states are connected by similar scattering vectors along $\Gamma$–M, yielding strong QPI peaks, but not along $\Gamma$–X (Fig.~\ref{fig:KS}d). But the nesting patterns evolve markedly near the Fermi level, where Kondo hybridization is strong. Two flat modes, $q_1^{+}$ and $q_1^{-}$, appear in both $\Gamma$–X and $\Gamma$–M directions. The integrated weight of $q_1^{\pm}$ is maximized above and below the Fermi level (Fig.~\ref{fig:QPI}c), with an energy separation consistent with the hybridization energy scale extracted from our Fano fits (Fig.~\ref{fig:fano}c). Meanwhile, $q_2$ and $q_3$ are reconstructed across the Kondo regime: $q_2$ narrows in momentum along $\Gamma$–M and also appears along $\Gamma$–X with a $\sqrt{2}$ geometric factor, as $q_3$ shifts to a lower momentum along $\Gamma$–M (Fig.~\ref{fig:QPI}a,b).

\ptitle{Model}
To interpret the QPI modes, we construct a tight-binding model that extends the tritelluride band structure to include Kondo hybridization (Fig.~\ref{fig:QPI}d; see full derivation in Methods). We start from an isolated Te layer, where the $p$-orbital geometry results in strong hopping parallel to the orbital lobes and weaker hopping perpendicular to them ($|t_\parallel| > |t_\perp|$, equation~(\ref{eq:TB_realspace_NN})). This anisotropy produces the familiar nested band structure of the RTe$_3$ family~\cite{yaoPRB2006}. Next, we account for the unit-cell doubling imposed by the UTe block, which halves the BZ and backfolds the bands~\cite{voitScience2000}. Although such back-folding was previously considered~\cite{brouetPRL2004}, the physical mechanism by which it enters the electronic potential was not, to the best of our knowledge, explicitly identified. Since the UTe block creates inequivalent electronic environments at the \textit{centers} of adjacent Te plaquettes (see $t_d$ terms in Fig.~\ref{fig:QPI}d for clarity), we model this term by introducing inequivalent next-nearest-neighbor hoppings along the Te1–U–Te1 and Te1–Te2–Te1 diagonals (equation~(\ref{eq:TB_realspace_NNN})). Finally, we include an effective Kondo coupling term ($t_\mathrm{K}$) between Te conduction orbitals and U $f$ orbitals (equation~(\ref{eq:HK_real})). Because the $f$ electrons reside on different lattice sites than the Te conduction electrons, the hybridization becomes momentum-dependent, and because U atoms occupy only half of the Te plaquettes, this coupling enhances the backfolding near \EF.

\ptitle{Model 2}
To compare our tight-binding model to our QPI data, we first diagonalize the Hamiltonian in momentum space to obtain the band structure. We present several band structure cuts in Fig.~\ref{fig:QPI}e,f (see Fig.~\ref{fig:QPI}g for context). To properly account for the influence of the BZ folding on the electronic structure, we project the spectral weights of all bands onto the appropriate momentum sector to obtain the experimentally relevant weight~\cite{voitScience2000} (see Methods). We then compute constant-energy spectral-weight maps in Fig.~\ref{fig:Energy_slices}a--d and auto-correlate these contours to simulate the corresponding QPI energy slices in Fig.~\ref{fig:Energy_slices}e--h. Finally, we plot the experimental QPI energy slices at the corresponding energies in Fig.~\ref{fig:Energy_slices}i--l (from the same data set as in Fig.~\ref{fig:QPI}a,b).

\ptitle{Comparing model to QPI}
By matching the simulated QPI slices in Fig.~\ref{fig:Energy_slices}e--h with the experimental data in Fig.~\ref{fig:Energy_slices}i--l, we can identify the origin of the scattering modes we observe. At negative energies, the nested band structure produces strong scattering intensity along $\Gamma$–M (Fig.~\ref{fig:QPI}e, Fig.~\ref{fig:Energy_slices}a). The avoided crossing between the original and backfolded bands splits this intensity, giving rise to the $q_2$ and $q_3$ modes. Entering the Kondo energy regime around \EF, the central square contour is reconstructed into a heavy band (Fig.~\ref{fig:Energy_slices}b)---scattering within this band produces the $q_1^{-}$ mode. At \EF, this central contour vanishes along with the $q_1$ mode (Fig.~\ref{fig:Energy_slices}c), destroying the Fermi-surface nesting typical of RTe$_3$ compounds. At positive energies, heavy quasiparticle bands form in the pockets between the original and backfolded contours (Fig.~\ref{fig:Energy_slices}d)---scattering within these pockets generates the $q_1^{+}$ mode. Scattering across opposite heavy-band pockets yields the renormalized $q_3$ mode (see also Fig.~\ref{fig:QPI}e). At these energies, spectral weight concentrates at the corners of the original contour; scattering between these corners results in a sharp $q_2$ mode along $\Gamma$–M and its $\sqrt{2}q_2$ counterpart along $\Gamma$–X (Fig.~\ref{fig:QPI}f, Fig.~\ref{fig:Energy_slices}d). We note that the diagonal-scattering mode $\sqrt{2}q_2$ exists throughout the entire energy range, but becomes more dominant once the central contour is destroyed and the spectral weight concentrates at these corners (Fig.~\ref{fig:Energy_slices}h), making it more visible experimentally (Fig.~\ref{fig:QPI}b, Fig.~\ref{fig:Energy_slices}l). To further compare our model with our experimental QPI results, we simulate energy-resolved QPI heat maps in Extended Data Fig.~\ref{fig:QPI_S}c,d, which reproduce the main features observed in Fig.~\ref{fig:QPI}a,b.

\ptitle{Speculations: CDW and magnetism}
Our tight-binding analysis shows that the QPI modes we observe are consistent with a Kondo-driven destruction of the nested Fermi surface. We now examine the consequences of this destruction for the correlated electronic phases of \but. In RTe$_3$ compounds, the nested Fermi surface drives CDW formation. Correspondingly, we observe no additional peaks indicative of a CDW in our \but\ STM topograph (Fig.~\ref{fig:KS}h) or STEM measurements (Extended Data Fig.~\ref{fig:STEM}). The destruction of nesting may also affect the magnetic ground state of the system. Strong Fermi-surface nesting promotes peaks in the Lindhard susceptibility at finite wavevectors, which favor antiferromagnetic RKKY interactions~\cite{ruPRB2008,aristovPRB1997}. When Kondo hybridization removes these nesting features, the finite-$q$ susceptibility is suppressed, possibly tilting the balance between antiferromagnetic and ferromagnetic exchange. Consistent with this picture, RTe$_3$ compounds order antiferromagnetically~\cite{ruPRB2008}, whereas \but\ is ferromagnetic~\cite{thomasArxiv2025}.

\ptitle{Speculations: Quantum criticality}
Tuning the balance between RKKY exchange and Kondo screening can drive quantum criticality and result in superconductivity~\cite{ranScience2019,paschenNatRPhys2021,checkelskyNatRMatt2024}, which has also been reported in several RTe$_3$ compounds under pressure~\cite{zoccoPRB2015}. While RKKY–Kondo criticality was not discussed in the context of the tritellurides, pressure should enhance the itinerant-$f$ electron coupling, analogous to the effect of extending the orbitals by replacing a rare-earth element with uranium. This raises the possibility that pressure tunes RTe$_3$ compounds toward quantum criticality, and that the superconductivity observed under pressure is related to heavy-fermion behavior. Furthermore, the CDW melts in the same pressure range where superconductivity emerges~\cite{zoccoPRB2015}, consistent with our observation that entering the Kondo regime destroys nesting and suppresses the CDW.

\ptitle{Conclusions and outlook}
Our results demonstrate that a heavy-fermion state can reshape correlated electronic orders beyond the well-known competition with an RKKY-mediated magnetic ground state. Using the tritelluride family as a platform, we demonstrate that entering the Kondo regime drives a transition from an RKKY-mediated antiferromagnetic phase—where nesting stabilizes CDW order—to a heavy-fermion state in which nesting is lost, the Fermi surface reconstructs into heavy quasiparticle bands, ferromagnetism emerges~\cite{thomasArxiv2025}, and CDW order disappears. We access the Kondo regime by broadening the studies of the 4$f$ RTe$3$ series to include the spatially-extended 5$f$ orbitals of actinides, identifying \but\ as a rare example of a two-dimensional heavy-fermion material~\cite{poseyNature2024,zhangSciAdv2018,broyles2025AdvMat}. Although our approach successfully drives the tritelluride system deep into the heavy-fermion regime, an open question is how the electronic ground state evolves across the RKKY–Kondo transition and whether quantum criticality emerges. Promising tuning knobs to explore this evolution include chemical doping, partial substitution on the $f$ sites, and exploiting the van der Waals structure to engineer interfaces between different tritellurides. Such control would enable direct tracking of how antiferromagnetism gives way to ferromagnetism, how CDW order melts as Kondo hybridization strengthens, and where superconductivity emerges within this competing landscape.

\newpage
\bibliography{refs}

\section*{Methods}
\subsection*{Crystal growth}
\noindent Single crystals of \but~were synthesized by the Molten Metal Flux method. The elements were mixed in the ratio U:Te = 1:15. The crucible containing the starting elements was sealed in a fused silica ampule and then heated up to 800$^\circ$C in a box furnace. The crystals grew as the temperature was reduced to 520$^\circ$C over 120 hours, after which the ampule was quickly removed from the furnace and the flux was decanted in a centrifuge.

\subsection*{STEM details}
\noindent Scanning transmission electron microscopy (STEM) and selected area electron diffraction (SAED) were performed on Thermo Fisher scientific (TFS) Themis Z G3 operated at 200 keV with Mel-build double tilt cryogenic air-free transfer holder. Annular dark-field (ADF-) STEM was operated with 18.9 mrad convergence semi-angle and SAEDs were collected with 850 nm diameter selected area aperture. The TEM samples were loaded into TEM air-free from an Argon filled glovebox using the air-free transfer TEM holder. Plan-view TEM samples were prepared by exfoliating \but\ flakes from a single crystal onto a polydimethylsiloxane (PDMS) gel stamp and mechanically transferring on to SiNx TEM grids for Norcada using a home-built transfer system inside the Argon filled glovebox. Cross-sectional TEM samples were prepared on TFS Helios 660 focus ion beam. Samples were kept in the glovebox to minimize degradation.

\subsection*{Quasiparticle interference analysis details}
\noindent To obtain the QPI maps used in Fig.~\ref{fig:QPI}a,b and Fig.~\ref{fig:Energy_slices}i--l, we first drift-corrected the spatial spectroscopic maps and then Fourier transformed each energy slice. Because \but\ is nearly tetragonal, we applied both mirror and fourfold symmetrization to the resulting $k$-space maps to improve signal-to-noise. To compute the heatmaps in Fig.~\ref{fig:QPI}a,b, we extracted finite-width line cuts along the $\Gamma$–X and $\Gamma$–M directions at each energy. We then stack these linecuts by energy and normalized each map at fixed $k$ so that the intensity summed over energy equals unity, which sharpens dispersive features.

\subsection*{Tight-binding model}
\noindent Our tight-binding model contains three coupling terms. The first is the nearest-neighbor (NN) hopping between Te $p$ orbitals, which is directionally anisotropic due to the orbital shapes. The second is the next-nearest-neighbor (NNN) hopping between diagonal Te sites. Because these diagonals pass above alternating atomic species located at the centers of the Te plaquettes (see Fig.~\ref{fig:QPI}d), the NNN hopping strength alternates from site to site, doubling the unit cell and generating back-folded bands. The third term is a Kondo-type hybridization between the Te $p$ states and the localized U~$f$ orbitals. Since the U atoms occupy the centers of every second plaquette, the hybridization also alternates between NN Te atoms and therefore further contributes to the same back-folding. Together, these three ingredients capture the essential electronic structure of \but: anisotropic Te conduction bands, BZ backfolding due to the doubled UTe unit cell, and heavy quasiparticle states emerging near the Fermi level. We construct our tight-binding Hamiltonian from these components and describe how the resulting electronic structure manifests experimentally.

\subsubsection*{Te-only model without orbital mixing.\ }
We begin with a square lattice of Te atoms and consider only the in-plane orbitals $p_x$ and $p_y$ located on each Te site $\mathbf r$. We introduce $c_{x,\mathbf r}^\dagger$ and $c_{y,\mathbf r}^\dagger$ as the creation operators for the Te $p_x$ and $p_y$ orbitals at site $\mathbf r$. The lattice vectors are taken as $\mathbf a_1=\hat{x}$ and $\mathbf a_2=\hat{y}$ for convenience. The NN hopping is directionally anisotropic: strong parallel to the orbital lobe direction (e.g.\ along $x$ for $p_x$) and weak in the perpendicular direction (e.g.\ along $y$ for $p_x$). It is this anisotropy that results in the familiar square-like contours of the tritelluride compounds~\cite{yaoPRB2006}. The real-space Hamiltonian is:
\begin{equation}
\begin{aligned}
H_{\text{NN}} &=
\sum_{\mathbf r} \Big[\epsilon_p \left(
c_{x,\mathbf r}^\dagger c_{x,\mathbf r}
+ c_{y,\mathbf r}^\dagger c_{y,\mathbf r}
\right) \\
&\qquad +
t_{\parallel}\, c_{x,\mathbf r}^\dagger c_{x,\mathbf r+\mathbf a_1}
+ t_{\perp}\, c_{x,\mathbf r}^\dagger c_{x,\mathbf r+\mathbf a_2} \\
&\qquad
+ t_{\perp}\, c_{y,\mathbf r}^\dagger c_{y,\mathbf r+\mathbf a_1}
+ t_{\parallel}\, c_{y,\mathbf r}^\dagger c_{y,\mathbf r+\mathbf a_2}
\Big] + \text{h.c.},
\end{aligned}
\label{eq:TB_realspace_NN}
\end{equation}
where $t_{\parallel}$ and $t_{\perp}$ are the NN hopping amplitudes parallel and perpendicular to the orbital lobe direction, and $\epsilon_p$ is the Te onsite energy.

Fourier transforming via $c_{\alpha,\mathbf r} = \frac{1}{\sqrt N}\sum_{\mathbf k} e^{i\mathbf k\cdot \mathbf r} c_{\alpha,\mathbf k}$ gives:
\begin{equation}
H^{(2\times2)}(\mathbf k) =
\begin{pmatrix}
h_{xx}(\mathbf k) & 0 \\
0 & h_{yy}(\mathbf k)
\end{pmatrix},
\end{equation}
where:
\begin{equation}
\begin{aligned}
h_{xx}(\mathbf k) &= \epsilon_p + 2t_{\parallel}\cos(\mathbf k\!\cdot\!\mathbf a_1)
                               + 2t_{\perp}\cos(\mathbf k\!\cdot\!\mathbf a_2),\\
h_{yy}(\mathbf k) &= \epsilon_p + 2t_{\perp}\cos(\mathbf k\!\cdot\!\mathbf a_1)
                               + 2t_{\parallel}\cos(\mathbf k\!\cdot\!\mathbf a_2).
\end{aligned}
\label{eq:hxx_hyy}
\end{equation}
\subsubsection*{Unit-cell doubling from the UTe layer.\ }
The UTe block beneath the Te layer doubles the unit cell, which halves the BZ and back-folds the electronic bands. While the simplest way to include such back-folding is by introducing an alternating on-site potential between NN Te atoms, this does not reflect the physical influence of the UTe block. Instead, the UTe block introduces an alternating environment \textit{between} adjacent centers of the Te plaquettes (see Fig.~\ref{fig:QPI}d). We therefore introduce the unit-cell doubling by considering alternating NNN hopping, which distinguishes between the Te1–U–Te1 and Te1–Te2–Te1 diagonal hopping along:
\begin{equation}
\mathbf d_1 = \mathbf a_1 + \mathbf a_2, \qquad
\mathbf d_2 = \mathbf a_1 - \mathbf a_2,
\end{equation}
and introduce anisotropic diagonal hopping terms as:
\begin{equation}
t_{\mathrm{Te}} = t_d - \delta t, \qquad
t_{\mathrm U}   = t_d + \delta t,
\end{equation}
where $t_d$ is the mean diagonal hopping and $\delta t$ quantifies the alternating anisotropy between Te1–Te2–Te1 and Te1–U–Te1 paths. To describe this alternating diagonal hopping, the hopping term along the $\mathbf{d_1}$ and $\mathbf{d_2}$ diagonals must alternate between $t_{\mathrm{Te}}$ and $t_{\mathrm{U}}$. We therefore introduce a modulation vector
$\mathbf Q=(\pi,\pi)$, such that the spatially-alternating diagonal hopping terms take the form of: $t_{\mathbf{d_1,d_2}}=t_d\pm\delta t_de^{i\mathbf Q\cdot\mathbf r}$. For diagonal hopping, the coupling between $p_x$ and $p_y$ orbitals is expected to be stronger than the same-orbital hopping, since the atoms are arranged diagonally with respect to one another (see Fig.~\ref{fig:QPI}d); this geometry introduces an orbital mixing term. We therefore introduce this NNN unit-cell doubling term into the Hamiltonian as:
\begin{equation}
\begin{aligned}
H_{\text{NNN}} =
\sum_{\mathbf r}\Big[
& (t_d + \delta t\, e^{i\mathbf Q\cdot\mathbf r})\,
c_{x,\mathbf r}^\dagger c_{y,\mathbf r+\mathbf d_1} \\
&+ (t_d - \delta t\, e^{i\mathbf Q\cdot\mathbf r})\,
c_{x,\mathbf r}^\dagger c_{y,\mathbf r+\mathbf d_2}
\Big] + \text{h.c.}
\end{aligned}
\label{eq:TB_realspace_NNN}
\end{equation}

Upon Fourier transformation, the Hamiltonian which combines both the NN and NNN terms, written in the basis:
\[
\Psi_{\mathbf k} = (c_{x,\mathbf k},\, c_{y,\mathbf k},\, c_{x,\mathbf k+\mathbf Q},\, c_{y,\mathbf k+\mathbf Q}),
\]
takes the form:
\begin{equation}
\begin{array}{l}
H^{(4\times4)}(\mathbf k)=\\[2pt]
\multicolumn{1}{c}{
\begin{pmatrix}
h_{xx}(\mathbf k) & h_{xy}(\mathbf k) & \alpha\, m(\mathbf k) & m(\mathbf k)\\
h_{xy}(\mathbf k) & h_{yy}(\mathbf k) & m(\mathbf k) & \alpha\, m(\mathbf k)\\
\alpha\, m(\mathbf k) & m(\mathbf k) & h_{xx}(\mathbf k+\mathbf Q) & h_{xy}(\mathbf k)\\
m(\mathbf k) & \alpha\, m(\mathbf k) & h_{xy}(\mathbf k) & h_{yy}(\mathbf k+\mathbf Q)
\end{pmatrix}.
}
\end{array}
\label{eq:H4x4}
\end{equation}
Here we also introduced a same-orbital NNN term similar to equation~(\ref{eq:TB_realspace_NNN}) with a prefactor $\alpha$, representing the weaker same-orbital back-folding coupling (e.g.\ $c_{x,\mathbf k}\!\leftrightarrow\!c_{x,\mathbf k+\mathbf Q}$) relative to the stronger mixed-orbital coupling (e.g.\ $c_{x,\mathbf k}\!\leftrightarrow\!c_{y,\mathbf k+\mathbf Q}$). We also defined:
\begin{equation}
\begin{aligned}
h_{xy}(\mathbf k) &= (t_{\mathrm U}+t_{\mathrm{Te}})\!
\left[\cos(\mathbf k\!\cdot\!\mathbf d_1)+\cos(\mathbf k\!\cdot\!\mathbf d_2)\right],\\
m(\mathbf k) &= (t_{\mathrm U}-t_{\mathrm{Te}})\!
\left[\cos(\mathbf k\!\cdot\!\mathbf d_1)-\cos(\mathbf k\!\cdot\!\mathbf d_2)\right],
\end{aligned}
\label{eq:hxy_m}
\end{equation}
and used $h_{xy}(\mathbf k+\mathbf Q)=h_{xy}(\mathbf k)$ and
$m(\mathbf k+\mathbf Q)=m(\mathbf k)$.

\subsubsection*{Kondo hybridization with U $f$ orbitals.\ }
The U atoms sit at the centers of every other Te plaquette
(see Fig.~\ref{fig:QPI}d) and hybridize
with the surrounding Te $p_x$ and $p_y$ orbitals.
Following the geometry of the square lattice, the relevant hybridization
paths connect each U site to Te atoms displaced by
$\pm \tfrac{1}{2}\mathbf d_1$ and $\pm \tfrac{1}{2}\mathbf d_2$.
To account for the fact that U atoms occupy only every second Te plaquette,
we introduce a periodic modulation
$s(\mathbf r) = s_0 \big(1 + e^{i\mathbf Q\cdot\mathbf r}\big)$,
where $\mathbf Q=(\pi,\pi)$ again describes the doubling of the unit cell. The real-space Kondo hybridization term is then written as:

\begin{equation}
\begin{aligned}
H_K = \;&
t_{K}\!\!\sum_{\mathbf r}\sum_{i=x,y}
s(\mathbf r)\,
f_{\mathbf r}^\dagger
\Big[
c_{i,\mathbf r+\frac{\mathbf d_1}{2}}
+ c_{i,\mathbf r-\frac{\mathbf d_1}{2}} 
\\
&\qquad\qquad+ c_{i,\mathbf r+\frac{\mathbf d_2}{2}}
+ c_{i,\mathbf r-\frac{\mathbf d_2}{2}}
\Big]
+ \text{h.c.}
\end{aligned}
\label{eq:HK_real}
\end{equation}
where $t_{K}$ is the effective Kondo coupling between
the Te $p$ orbitals and the U~$f$ orbital, and $f_{\mathbf r}^\dagger\!$ is the creation operator for the $f$ electrons. Fourier transforming yields:

\begin{equation}
\begin{aligned}
H_K =  \tilde t_{K}
\sum_{\mathbf k}
\sum_{i=x,y}
f_{\mathbf k}^\dagger
\Big[
& g(\mathbf k)\, c_{i,\mathbf k}
\\&+ g(\mathbf k+\mathbf Q)\, c_{i,\mathbf k+\mathbf Q}
\Big]
+ \text{h.c.}
\end{aligned}
\label{eq:HK_kspace}
\end{equation}
where $\tilde t_{K}=4s_0 t_{K}$ absorbs the overall normalization, and we defined the hybridization form factor:
\begin{equation}
\begin{aligned}
g(\mathbf k) &=
\cos\!\left(\tfrac{1}{2}\mathbf k\!\cdot\!\mathbf a_1\right)
\cos\!\left(\tfrac{1}{2}\mathbf k\!\cdot\!\mathbf a_2\right) \\
&\qquad= \tfrac{1}{2}\!\left[
\cos\!\left(\tfrac{1}{2}\mathbf k\!\cdot\!\mathbf d_1\right)
+ \cos\!\left(\tfrac{1}{2}\mathbf k\!\cdot\!\mathbf d_2\right)
\right].
\end{aligned}
\label{eq:gk}
\end{equation}
We note that Fourier transforming yields an additional term which couples $f_{\mathbf k+\mathbf Q}$ to the Hamiltonian identically to how $f_{\mathbf k}$ couples into the Hamiltonian in equation~(\ref{eq:HK_kspace}). This additional redundant term is a result of the mathematical trick we used in equation~(\ref{eq:HK_real}) to introduce the modulating Kondo site by defining the periodic modulation $s(\mathbf{r})$. To avoid redundancy, we keep only the term in equation~(\ref{eq:HK_kspace}). Alternatively, one can define new operators $f_{\mathbf \pm}=\frac{1}{\sqrt{2}}(f_{\mathbf k}\pm f_{\mathbf k+Q})$, resulting in the same $5\times5$ Hamiltonian structure below, and an extra redundant state which is decoupled from the Hamiltonian.

\subsubsection*{Full folded Hamiltonian with Kondo coupling.\ }
Combining equations~(\ref{eq:H4x4}) and~(\ref{eq:HK_kspace}), the complete Hamiltonian in the basis:
\begin{equation}
\Psi_{\mathbf k}
= (c_{x,\mathbf k},\, c_{y,\mathbf k},\, c_{x,\mathbf k+\mathbf Q},\, c_{y,\mathbf k+\mathbf Q},\, f_{\mathbf k})
\label{eq:basis5}
\end{equation}
takes the form:
\par\bigskip
\begin{widetext}
\begin{equation}
H^{(5\times5)}(\mathbf k) =
\begin{pmatrix}
h_{xx}(\mathbf k) & h_{xy}(\mathbf k) & \alpha\, m(\mathbf k) & m(\mathbf k) & \tilde t_{K} g(\mathbf k)\\
h_{xy}(\mathbf k) & h_{yy}(\mathbf k) & m(\mathbf k) & \alpha\, m(\mathbf k) & \tilde t_{K} g(\mathbf k)\\
\alpha\, m(\mathbf k) & m(\mathbf k) & h_{xx}(\mathbf k+\mathbf Q) & h_{xy}(\mathbf k) & \tilde t_{K} g(\mathbf k+\mathbf Q)\\
m(\mathbf k) & \alpha\, m(\mathbf k) & h_{xy}(\mathbf k) & h_{yy}(\mathbf k+\mathbf Q) & \tilde t_{K} g(\mathbf k+\mathbf Q)\\
\tilde t_{K} g(\mathbf k) & \tilde t_{K} g(\mathbf k) &
\tilde t_{K} g(\mathbf k+\mathbf Q) & \tilde t_{K} g(\mathbf k+\mathbf Q) & \epsilon_f
\end{pmatrix}.
\label{eq:H5x5}
\end{equation}
\end{widetext}
Here, $\epsilon_f$ denotes the effective on-site energy of the U~$f$ level.  
This Hamiltonian captures (i)~the anisotropic Te $p_x$–$p_y$ conduction
bands, (ii)~the $\mathbf Q=(\pi,\pi)$ unit-cell doubling induced by the
UTe layer, and (iii)~the Kondo hybridization between the Te $p$ states
and the U~$f$ orbital.

\subsubsection*{Spectral weight and projection procedure.\ }
When crystals host several competing periodic potentials---such as
antiferromagnetic order, charge density waves, or, in our case---the
distinct periodicities of the Te and UTe layers---the resulting
Hamiltonian couples electronic states at $\mathbf k$ and
$\mathbf k+\mathbf Q$, where $\mathbf Q$ is the wavevector of the
additional periodicity.  Consequently, the basis of
equation~(\ref{eq:basis5}) contains both the original $\mathbf k$ operators
$(c_{x,\mathbf k},\, c_{y,\mathbf k})$ and the folded
$\mathbf k+\mathbf Q$ operators
$(c_{x,\mathbf k+\mathbf Q},\, c_{y,\mathbf k+\mathbf Q})$. Accordingly, when we diagonalize equation~(\ref{eq:H5x5}):
\begin{equation}
H^{(5\times5)}(\mathbf k)\,|\psi_n(\mathbf k)\rangle
= E_n(\mathbf k)\,|\psi_n(\mathbf k)\rangle
\label{eq:eig5a}
\end{equation}
we obtain five bands: $E_n(\mathbf k)$, and corresponding \textit{mixed} wavefunctions: $\psi_n(\mathbf k)$. Namely, these wavefunctions are not pure Bloch states with a single
crystal momentum.  Instead, the wavefunctions are a superposition of the
five basis states listed in equation~(\ref{eq:basis5}). To calculate the experimentally observed spectral weight, we project these wavefunctions to the original $\mathbf k$ orbitals~\cite{voitScience2000}. We do this by writing the wavefunctions as:
\begin{equation}
|\psi_n(\mathbf k)\rangle
= \sum_{\alpha=1}^{5} u_{\alpha n}(\mathbf k)\,|\alpha\rangle ,
\label{eq:u_def}
\end{equation}
where $\alpha=1,2,3,4,5$ label the basis orbitals
$(c_{x,\mathbf k},\, c_{y,\mathbf k},\, c_{x,\mathbf k+\mathbf Q},\, c_{y,\mathbf k+\mathbf Q},\, f_{\mathbf k})$, and then calculating the projected components $\alpha=1,2$ for the Te $p$ states, and
$\alpha=5$ for the U~$f$ state:
\begin{align}
W_n^{(p)}(\mathbf k)
&= |u_{1n}(\mathbf k)|^{2} + |u_{2n}(\mathbf k)|^{2}, \\[2pt]
W_n^{(f)}(\mathbf k)
&= |u_{5n}(\mathbf k)|^{2}.
\end{align}
The observable spectral function is then obtained by weighting each band
with its orbital content and applying Lorentzian broadening,
\begin{equation}
\begin{aligned}
A^{(p,f)}(\mathbf k,\omega)
&=
\\&\frac{1}{\pi}\sum_{n}
W_n^{(p,f)}(\mathbf k)\,
\frac{\delta}{(\omega - E_{n}(\mathbf k))^{2} + \delta^{2}} .
\end{aligned}
\label{eq:Akw_final}
\end{equation}
where $\delta$ represents phenomenological lifetime broadening.

Since the weight of the backfolded $\mathbf k+\mathbf Q$ sector ($\alpha=3,4$) manifests experimentally by mixing with the original $\mathbf k$ sector, it is strongest where the original and folded bands intersect, and mixing is maximal. In our model, the alternating NNN hopping enhances this mixing near the boundary of the reduced BZ. The Kondo hybridization further increases this mixing near the Fermi level.

\subsubsection*{QPI calculation from orbital-resolved autocorrelations.\ }
To obtain the QPI patterns associated with
the Te conduction states and the U~$f$ states, we compute the
autocorrelations of the corresponding spectral functions
$A^{(p)}(\mathbf k,\omega)$ and $A^{(f)}(\mathbf k,\omega)$ defined in
equation~(\ref{eq:Akw_final}).  For each energy $\omega$, the
orbital-resolved QPI intensities are:
\begin{equation}
I^{(p,f)}(\mathbf q,\omega)
= \sum_{\mathbf k}
A^{(p,f)}(\mathbf k,\omega)\,
A^{(p,f)}(\mathbf k+\mathbf q,\omega).
\label{eq:autocorr_basic}
\end{equation}
Because tunneling into the U~$f$ orbitals is weaker than into the Te $p$
orbitals, and because the $f$ orbitals are more spatially localized, we weight the two channels differently and include
an additional momentum-space damping in the $f$ contribution.  The resulting QPI
intensity is:
\begin{equation}
\begin{aligned}
I_{\text{QPI}}(\mathbf q,\omega)
&=
\\&|t_p|^{2}\,I^{(p)}(\mathbf q,\omega)
+ |t_f|^{2}\,e^{-|\mathbf q|/\kappa}\,I^{(f)}(\mathbf q,\omega),
\end{aligned}
\label{eq:qpi_combined}
\end{equation}
where, $t_p$ and $t_f$ set the relative tunneling amplitudes, and $\kappa$
controls the short-range nature of the $f$ channel. For simplicity, we neglect cross-scattering between the $p$ and $f$ states. 

\subsubsection*{Comparing to experiment.\ }
Using the procedure outlined above, we calculate the tight-binding model and simulate QPI using the following parameters.

For the Te on-site energy and nearest-neighbor hoppings,
\[
\epsilon_{p} = -1.8~\mathrm{eV}, \qquad
t_{\parallel} = 3.5~\mathrm{eV}, \qquad
t_{\perp} = -0.7~\mathrm{eV}.
\]

For the diagonal next-nearest-neighbor amplitudes associated with unit-cell doubling,
\[
t_{d} = 0.05~\mathrm{eV}, \qquad
\delta t = -0.15~\mathrm{eV},
\]
with back-folded orbital anisotropy $\alpha = 0.2$.

For the Kondo hybridization and effective $f$-orbital on-site energy,
\[
\tilde t_{K} = 0.11~\mathrm{eV}, \qquad
\epsilon_{f} = 0.02~\mathrm{eV}.
\]

Throughout all calculated spectral weights and QPI simulations, we use lifetime broadening, $\delta = 0.01~\mathrm{eV}$ and apply a factor of $0.01$ to the $f$-orbital spectral weight to account for the diminished tunneling into this channel ($|t_p|^{2}=1$, $|t_f|^{2}=0.01$). In the QPI simulations, we also include momentum-space damping of the $f$ contribution with $\kappa=\pi/2$.

To further compare the model with the experimental QPI data, we simulate QPI scattering maps over the full experimental energy range. From these simulations, we extract linecut intensities along the $\Gamma$–M and $\Gamma$–X directions and assemble energy-resolved directional heatmaps (Extended Data Fig.~\ref{fig:QPI_S}c,d). As in Fig.~\ref{fig:QPI}a,b (reproduced in Extended Data Fig.~\ref{fig:QPI_S}a,b for clarity), the heatmaps are normalized by momentum such that the sum over energy at each momentum equals unity. To simulate experimental conditions, random noise is added, and the linecuts are obtained by integrating over a finite momentum width.

\section*{Acknowledgments}
\noindent We thank Harris Pirie for useful discussions. G.T. was supported by the Israeli Council of Higher Education Quantum Technology Fellowship. D.K.M. acknowledges support by the U.S.\ Department of Energy, Office of Science, Basic Energy Sciences, under Award No.\ DE-FG02-05ER46225. C.G., S.H.S., and I.E. were supported by the Rowland Institute at Harvard. This work was carried out in part through the use of Massachusetts Institute of Technology (MIT) nano facilities. This work was performed in part at the Harvard University Center for Nanoscale Systems (CNS); a member of the National Nanotechnology Coordinated Infrastructure Network (NNCI), which is supported by the National Science Foundation under NSF Award No.\ ECCS-2025158.

\section*{Author contributions}
\noindent G.T. and R.K. contributed equally to this work. J.E.H supervised the project.  G.T., R.K., Y. C., Y.W., and M.L. performed the STM experiments and analyzed the data. G.T., D.G., D.K.M., B.Y., and J.E.H. developed the theoretical model. C.G., S.H.S., and I.E.B. performed the STEM experiments. J.S., J.Z, D.M., and S.R prepared the samples. G.T. and J.E.H. wrote the paper with input from all authors.
\pagebreak
\newpage

\setcounter{figure}{0}
\setcounter{table}{0}
\renewcommand{\thetable}{\arabic{table}}
\makeatletter
\renewcommand*{\fnum@figure}{\textbf{Extended Data Fig.~\thefigure}\,}
\renewcommand*{\@caption@fignum@sep}{\textbf{\,\textbar\,}}
\renewcommand*{\fnum@table}{\textbf{Extended Data Table \thetable}\,}
\makeatother

\begin{figure*}
    \centering
    \includegraphics[width=\linewidth]{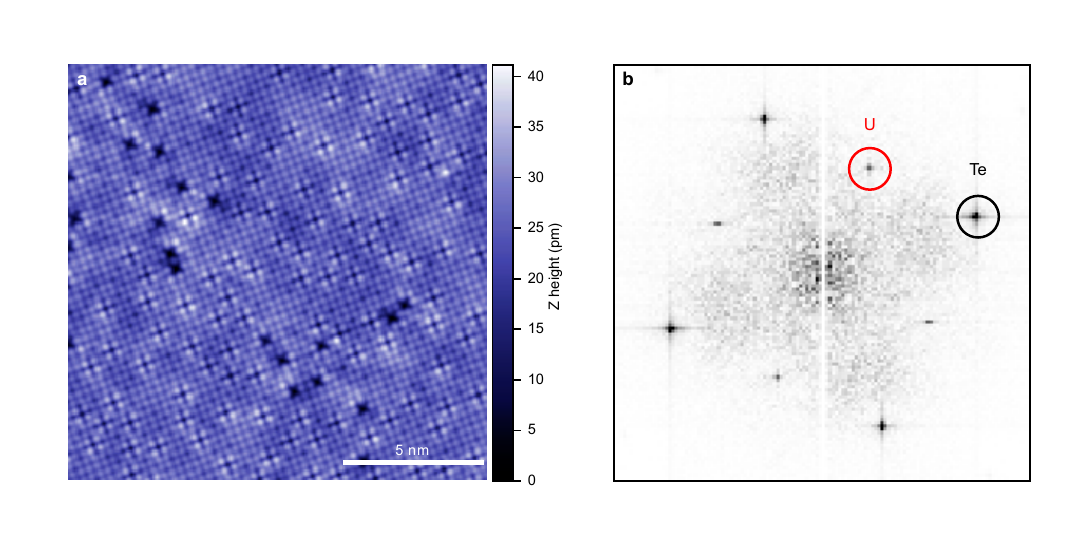}
    \caption{\textbf{STM topograph of the surface of \but.}
    \textbf{a}, Large field of view of the \but\ surface topograph. A crop from this scan was used in Fig.~\ref{fig:KS}g. \textbf{b}, Fourier transformation of (\textbf{a}). The Bragg peaks corresponding to the U and Te periodicities are marked. STM topograph was measured at $T = 4.5$ K with sample bias $\Vs = -200$ mV and current setpoint $\Is = 100$ pA.}
    \label{fig:full_topo}
\end{figure*}

\begin{figure*}[ht]
    \centering
    \includegraphics[width=\linewidth]{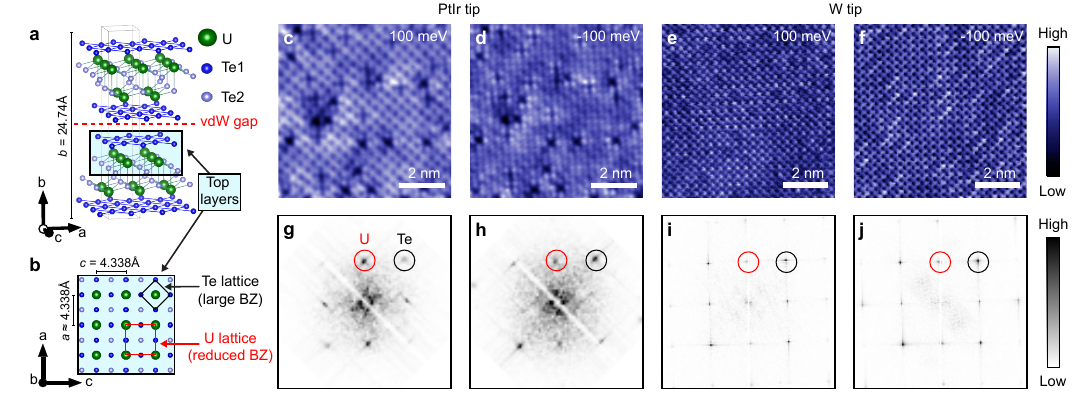}
    \caption{\textbf{Effect of bias and tip material on STM topography.} \textbf{a}, Crystal structure of \but, showing the vdW gap and the layers exposed by cleaving.
\textbf{b}, Top view of the \but\ structure, highlighting the Te and U sub-lattices.
\textbf{c},\textbf{d}, STM topographs of the \but\ surface acquired with a PtIr tip (as used for all main-text data). 
\textbf{e},\textbf{f}, STM topographs of the \but\ surface acquired with a W tip. For each tip, scans were taken on the same area at $V_\mathrm{sample}=100$ mV and $V_\mathrm{sample}=-100$ mV.
\textbf{g}--\textbf{j}, Corresponding Fourier transformations of the topographs in (\textbf{c}--\textbf{f}). Bragg peaks from the Te periodicity (black circles) and the U periodicity (red circles) are visible. With the PtIr tip, the U peaks dominate at positive bias, while the Te peaks dominate at negative bias. 
This behavior matches the main text: in Fig.~\ref{fig:KS}g we use negative bias and observe that the Te lattice corrugation dominates the topographs, whereas in Fig.~\ref{fig:fano}a we use positive bias and the U lattice corrugation dominates. Since the scans shown in (\textbf{c}) and (\textbf{d}) were taken on the same area, we confirm that the bias dependence of the Bragg peaks originates from an electronic effect rather than spatial variations across the sample. In contrast to the PtIr tip measurements (\textbf{g},\textbf{h}), both Te and U Bragg peaks are clearly resolved in the W-tip Fourier transformations (\textbf{i},\textbf{j}) with no significant bias dependence, indicating that the tunneling into the different sub-lattices depends on the tip material in addition to the sample bias.
Measurements were performed under the following conditions and setpoints:
(\textbf{c},\textbf{d}) $T=5$ K, $\Is=200$ pA;
(\textbf{e},\textbf{f}) $T=10.3$ K, $\Is=150$ pA.}
    \label{fig:topo_tips}
\end{figure*}

\begin{figure*}[ht]
    \centering
    \includegraphics[width=\linewidth]{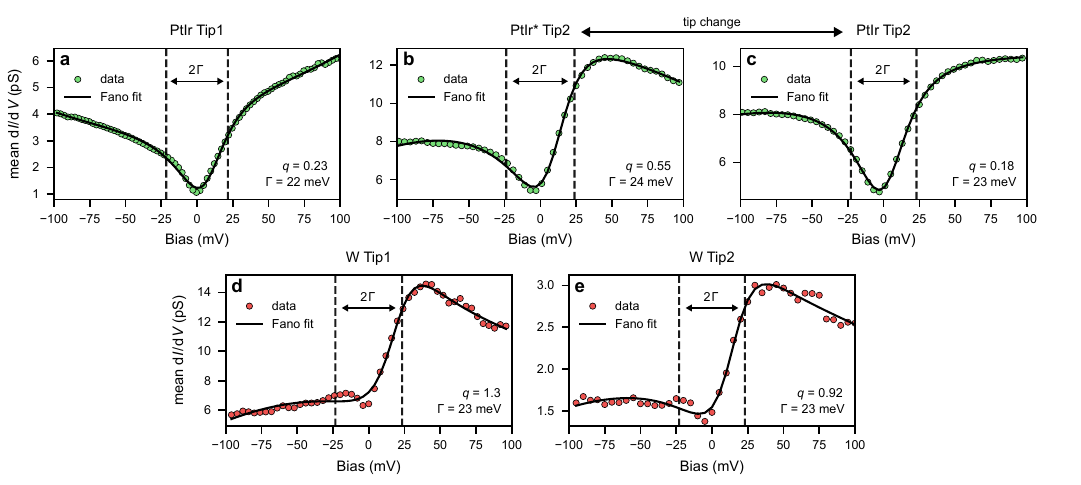}
    \caption{\textbf{\but\ spectra acquired with different tips show different Fano curves.} In all panels, colored points show data averaged over several spectra, and black curves show Fano fits as described in the main text. The measurements span four different samples and multiple tips.
\textbf{a}, Spectra obtained with a PtIr tip, corresponding to the averaged data from the map shown in Fig.~\ref{fig:fano}.
\textbf{b},\textbf{c}, Sequential spectra on another sample using a second PtIr tip. Both (\textbf{b}) and (\textbf{c}) were taken in the same area, but the tip changed during acquisition, producing significantly larger $q$ in (\textbf{b}) compared to (\textbf{c}). Since the spectrum in (\textbf{c}) resembles those obtained in other measurements, such as in (\textbf{a}), we conjecture that in (\textbf{b}) the tip picked up a surface atom that altered the tip apex and that in (\textbf{c}) this atom dropped. 
\textbf{d},\textbf{e}, Spectra from two additional samples using two different W tips, both displaying clear Fano behavior with a $q$ parameter significantly larger than any measurement using a PtIr tip. In all panels, the energy span $2\Gamma$ is indicated by the dashed vertical lines and black arrow. Despite variations in $q$, the extracted hybridization width $\Gamma$ is statistically the same across all datasets from (\textbf{a}) through (\textbf{e}). 
We interpret this behavior by noting that $q$ reflects the tunneling asymmetry between tunneling into the conduction electrons and tunneling into the $f$ electrons. The tip material can change the tunneling amplitude into each electronic channel, and thus their ratio, which is represented by $q$. In contrast, $\Gamma$ reflects an intrinsic property of the sample---the Kondo energy scale of \but---and should be independent of tip material, consistent with our observations. We therefore conclude that the observed variations in spectroscopy and topography reflect changes in the tunneling amplitudes into the $f$ and conduction states induced by variations in the tip material and sample bias. Measurements were performed under the following conditions and setpoints:
(\textbf{a}) $T=7.2$ K, $\Vs=100$ mV, $\Is=150$ pA;
(\textbf{b},\textbf{c}) $T=7.1$ K, $\Vs=-200$ mV, $\Is=400$ pA;
\textbf{(d)} $T=10.3$ K, $\Vs=-100$ mV, $\Is=150$ pA;
\textbf{(e)} $T=10.1$ K, $\Vs=150$ mV, $\Is=100$ pA.
All spectra were acquired with a lock-in modulation amplitude of $\Ve=4$ mV.}
    \label{fig:fano_tips}
\end{figure*}

\begin{figure*}
    \centering
    \includegraphics[width=\linewidth]{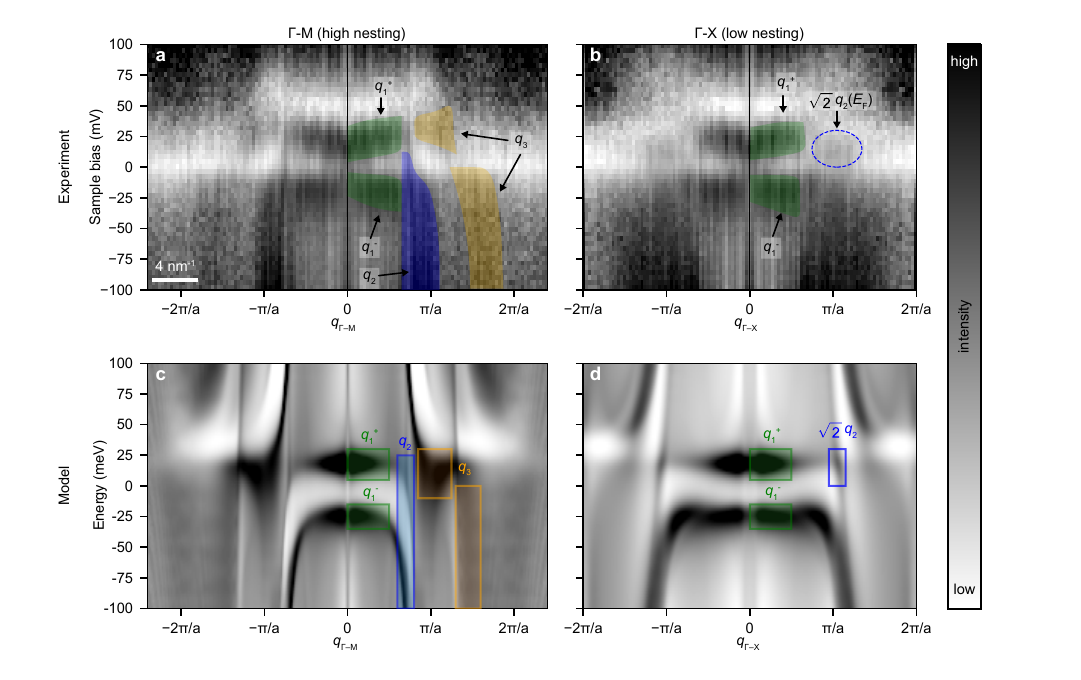}
    \caption{\textbf{Comparison between experimental and simulated QPI heatmaps.} \textbf{a},\textbf{b}, Reproduction of the experimental QPI heatmaps from Fig.~\ref{fig:QPI}. \textbf{c},\textbf{d}, Simulated QPI intensities calculated using the model described in Methods. The dominant scattering modes observed experimentally are indicated in the simulated data.}
    \label{fig:QPI_S}
\end{figure*}

\begin{figure*}
    \centering
    \includegraphics[width=\linewidth]{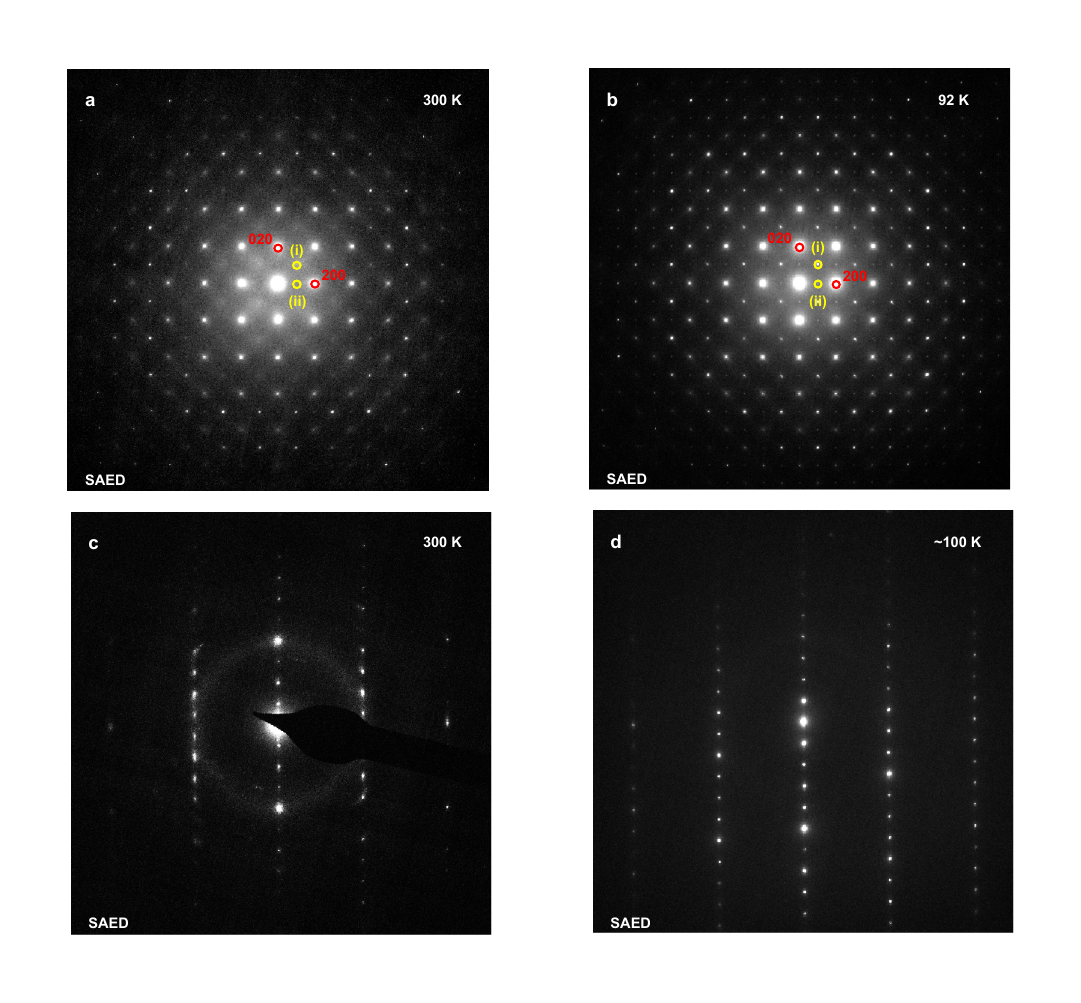}
    \caption{\textbf{STEM measurements show no CDW.}
    \textbf{a}--\textbf{d}, Combined in-plane (\textbf{a},\textbf{b}) and cross-sectional (\textbf{c},\textbf{d}) selected-area electron diffraction (SAED) patterns of \but\ collected at room temperature and liquid nitrogen temperatures. In the in-plane geometry, cooling leads to sharpening of the 110 peaks (i) and the appearance of weak 100 peak (ii), consistent with changes in stacking order. The cross-sectional SAED patterns show no changes in peak positions or intensities upon cooling. No charge-order superlattice peaks were observed in either the in-plane or out-of-plane orientations.}
    \label{fig:STEM}
\end{figure*}

\end{document}